\documentclass[twocolumn,letterpaper,english]{revtex4}
\usepackage[T1]{fontenc}
\usepackage[latin1]{inputenc}
\usepackage{graphicx}

\makeatletter

\usepackage{graphicx}

\usepackage{graphicx}

\usepackage{babel}

\usepackage{babel}
\def\etal{{\it et\thinspace al.}\ }
\def\ka{{K$\alpha$}\ }
\def\kb{{K$\beta$}\ }
\def\ra{{$\rightarrow$}\ }
\makeatother
\begin{document}

\title{Pumping \ka Resonance Fluorescence by Monochromatic X-Ray Sources} 

\author{Sultana N. Nahar$^{1,\ast}$ and Anil K.
Pradhan$^{1,2,3,\dagger}$} 

\affiliation{$^{1}$Department of Astronomy,$^{2}$ Chemical Physics Program,
$^3$ Biophysics Graduate Program, The Ohio State University, Columbus, Ohio
43210.}

\begin{abstract}

 We demonstrate the correspondence between theoretically calculated
K-shell resonances lying below the K-edge in multiple ionization states
of an element \cite{pra09}, 
and recently observed \ka
resonances in high-intensity X-ray free-electron laser (XFEL)
 plasmas \cite{vin12}. 
Resonant absorptions in aluminum ions are computed and found to
reproduce experimentally observed features.
Results are also presented for titanium for possible observation of \ka
resonances in the 4.5-5.0 keV energy range. 
A possibly sustainable excitation mechanism for \ka resonance 
fluorescence might be implemented using two
monochromatic X-ray beams tuned to the K-edge and the \ka
resonant energies simultaneously. This targeted ionization/excitation
would create inner-shell vacancies via Auger decay, as well as pump \ka
resonances. The required X-ray fluence to achieve resonance fluorescence
would evidently be much less than in the XFEL
experiments, and might enable novel biomedical applications.

\textbf{~}
\textbf{PACS number(s)} : 32.80.Aa, 32.80.Hd, 61.80.Cb, 87.53.-j,
61.80.Cb
\end{abstract}
\maketitle

 Excitation of resonances in atoms with inner-shell vacancies has been
under considerable theoretical and experimental study recently
\cite{pra09,vin12,mon09,kan11,roh12}.
In addition to basic phenomena of intrinsic physical interest, 
there are two main reasons for these
studies: the potential for novel practical applications in principle
\cite{pra09,mon09}, and the
advent of high-intensity sources, such as the X-ray free-electron laser
(XFEL) \cite{you10}, required to create a high-energy-density 
(HED) \cite{drake} plasma containing the "hollow" atomic ions.
In earlier theoretical investigations of K-shell resonances 
[1,3] we showed
that the cross sections for resonant photo-excitations of \ka, \kb,
etc. are orders of magnitude
higher than the photoionization continua lying {\it below} the K-edge.
The corresponding resonances or channels [3] become energetically accessible 
following K-shell ionization and Auger decays that open up mutliple
electronic vacancies in higher shells, particularly in high-Z atoms such
as iron, platinum and gold \cite{pra09,nah11}. Following works 
have labelled these as "hidden" resonances [4] created in
double-core-excited 'hollow' ions [5].

 However, there are several problems in probing the dynamics of
inner-shell resonant excitation. A monochromatic source is
necessary to scan across the resonance energies. Sufficiently high
intensities are necessary to create and achieve plasma
conditions for deep electronic shell vacancies to exist. Extremely
short pulse time-scales comparable to intrinsic atomic 
transition rates are required for {\it in situ} studies of resonant
excitation. These criteria and problems have been addressed by 
recent XFEL studies,
albeit for low Z atomic species where resonant rates are
much smaller than the high Z atoms considered theoretically
\cite{pra09,mon09}. 
Neon and aluminum ions were the focus of such experiments
at the Stanford Linear Accelerator (SLAC) using the 
Linac Coherent Light Source (LCLS) XFEL \cite{roh12,vin12}. 
The peak intensities are extremely high, approximately
10$^{17}$ W cm$^{-2}$. Experimental diagnosis
centered on \ka resonance fluorescence (hereafter RFL): K-shell ionization
followed by \ka emission creating a vacancy in the L-shell, following by
resonance K \ra L resonant excitation. Earlier theoretical
calculations for transition probabilities and cross sections of several
high Z elements
\cite{nah08,sur08,nah11} were carried out for all ionization states possibly
involved in \ka excitations, from H-like to F-like ions, and all
resonances from \ka to K$\eta$, i.e. K \ra L, M, N, O, P
\cite{pra09,nah11}. In this
{\it Letter} we establish a correspondence between the observed \ka
resonances in Al ions \cite{vin12} and the computed 
resonant absorption that drives \ka RFL, 
leading to an "enhancement" of the Auger effect (Fig.2, Ref. [1]).

 The extremely high intensities needed to first create and then pump
these resonances appear to make it impractical for any potential
applications in the near future. While X-ray sources such
as synchrotrons \cite{elleaume,lesech}, 
and now XFEL, are monochromatic and capable of high
fluence that enables new experimentation, 
they are not readily available or suitable for technological or
biomedical use. Whereas the task of creating deep inner-shell
holes remains daunting in ordinary situations, it is worth
asking what might be the {\it least} energetic requirement to pump RFL
efficiently. In this report we also propose a schematic
twin-beam monochromatic X-ray setup which
might, in turn, enable applications such as localized X-ray deposition
using high-Z radiosensitzation in radiation therapy.
Other effects related to Auger transitions, such as Rabi
oscillations, and ultra-short femtosecond monochromatic X-ray pulses,
are briefly discussed. 

 The LCLS-XFEL is sufficiently intense to create a solid-density plasma,
where several ionization states of an element may exist with K-shell
vacancies. When the FEL energy equals the \ka energy of an ion, RFL
occurs and manifests itself as \ka emission which can be detected.
The aluminium plasma created in the LCLS-XFEL experiment exhibited RFL
for several ions at energies below the
K-edge [2]. The resonance energies and strengths may be computed 
for all such ions, and all resonance transitions
from the K-shell upwards leading up to K-edge [1]. However, including
fine structure, there are a
large number of \ka transitions that come into play: a total of 112
transitions are quantum mechanically allowed for
all ions from H- to F-like \cite{nah03,nah08,aas}. 
Many of these overlap among adjacent
ionization states. In order to facililate a correspondence with
\ka RFL measurements, we have computed only the \ka resonant absorption
attenuation coefficients (cm$^2$/g) taking account of overlapping profiles. The
resulting \ka absorption that drives RFL pumping is shown in Fig. 1, and
compared with the experimental results.

\begin{figure}
%\begin{center}
%\includegraphics[width=95mm,height=75mm]{pxacres-al.eps}
\includegraphics[width=95mm,height=75mm]{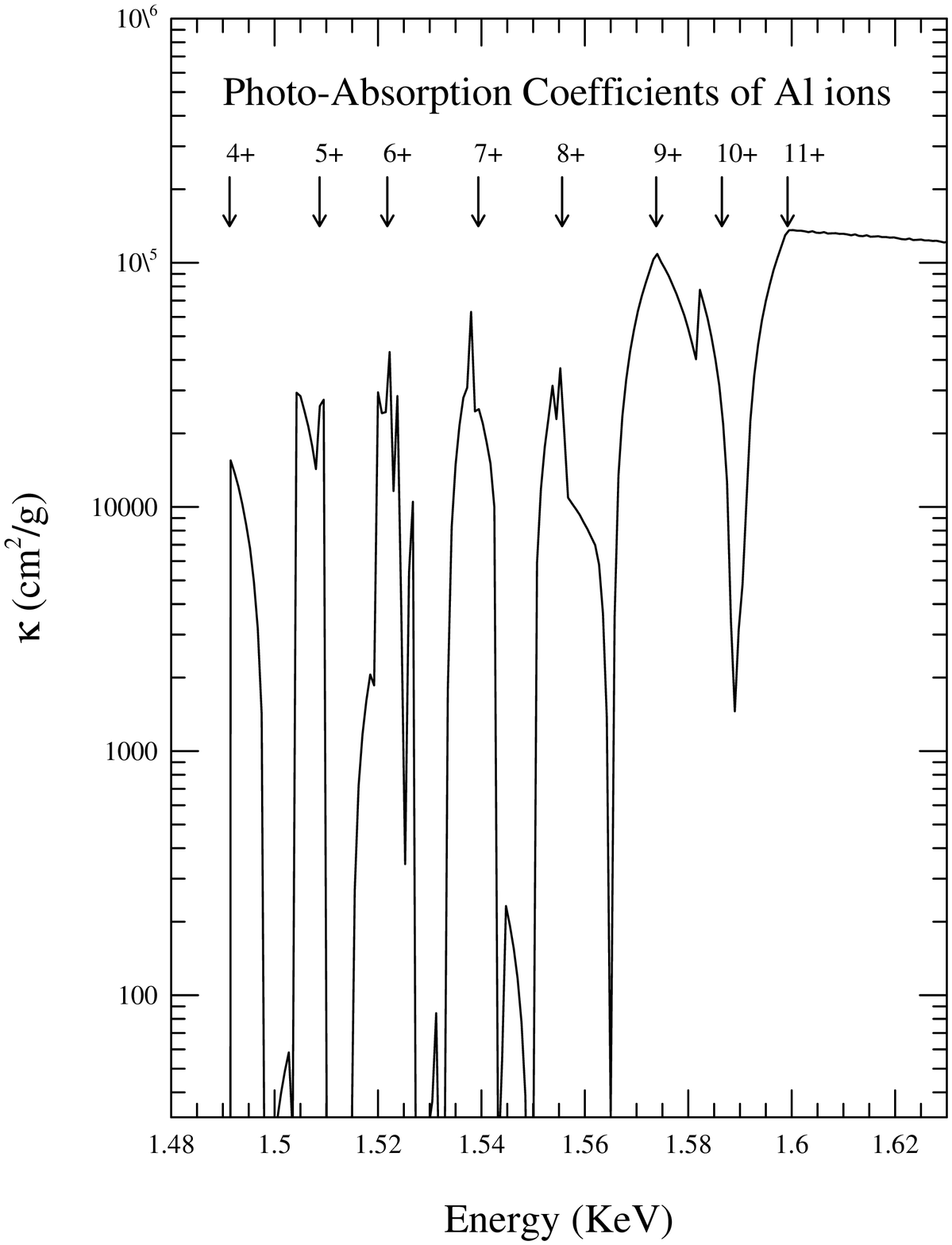}
\includegraphics[width=75mm,height=60mm]{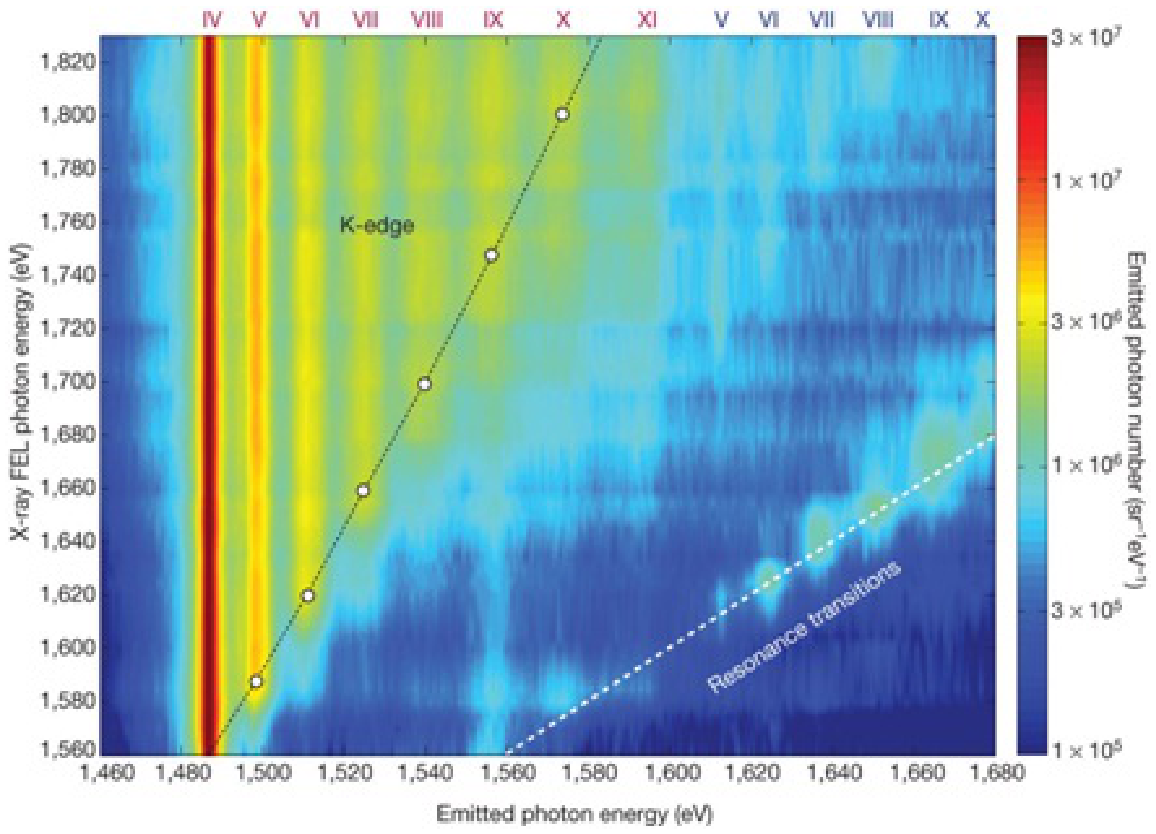}
%\end{center}
\caption{(color online). 
Top: \ka resonance fluorescence in 
aluminum plasma: Theoretically computed \ka absroption features in Al ions 
isolectronic with fluorine to helium are shown (ionization states are labeled
according to ion charge in the upper panel and roman numerals in the
lower panel). All contributing \ka transition
strengths are added together and 
shown in terms of X-ray attenuation coefficients $\kappa$
(cm$^2$/g) through the Al plasma (convolved over a gaussian FWHM of 10
eV). Bottom: Experimental
measurements from the LCLS-XFEL [2] (reproduced with permission). 
The dashed line on the right shows
the resonance transtions with increasing \ka emission intensity
corresponding to the theoretical absorption complexes. The observed \ka
features occur when the XFEL energy equals the emitted photon energy.
 \label{fig:alres}}
\end{figure}

 The cross sections used to compute the resonance structures in
Fig.~\ref{fig:alres} were convolved with a small gaussian beam width of
10 eV FWHM. The calculations are carried out using the Breit-Pauli
version of the atomic structure and R-matrix codes \cite{aas}.
Evidently, the \ka absorption resonance complexes for each ion can be associated
with the \ka emission seen experimentally (lower panel). The rising trend in 
relative intensities, measured as "Emitted photon number (sr$^{-1}$eV$^{-1}$)"
[2], is also evident in the theoretical \ka resonance strengths for 
successively higher ionization states of Al (upper panel). The experimental data
in the lower panel exhibit overlaps and different total intensities of
the various \ka complexes. That is also seen qualitatively in the top
panel with aggregate \ka resonance strengths for each ion.
 A more quantitiative description of averaged energies
and \ka resonance strengths is given in Table~1 for H- to F-like 
ionization states for aluminum. The number of transitions, and relative 
positions and 
\ka strengths, can be discerned readily from the fine structure averaged
energies and cross sections (megabarns) derived from 
resonance oscillator strengths.

\begin{figure}
%\begin{center}
%\includegraphics[width=95mm,height=75mm]{pxacres-ti.eps}
\includegraphics[width=95mm,height=75mm]{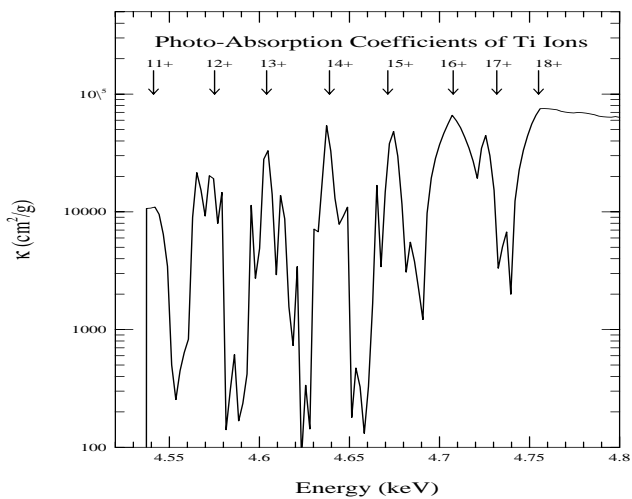}
%\end{center}
\caption{Titanium \ka resonance complexes, as in Fig.~\ref{fig:alres}
(top panel for aluminum ions). 
For clarity, the H-like Ti$^{19+}$ \ka resonances are omitted since
they lie significantly above the He-like Ti$^{18+}$ complex (also in
Fig.~\ref{fig:alres} for Al ions). The Ti \ka resonance fluorescence has
not yet been observed, but the energies are in the accessible range of
beam energies at the LCLS-XFEL. The XFEL energies may be scanned across
the calculated average \ka energies for each ion given in Table~1. 
\label{fig:tires}}
\end{figure}

 Table~1 also presents the positions and \ka resonance strengths for
possible experimental detection thereof for titanium. 
Fig.~\ref{fig:tires} shows the \ka resonance complexes for titanium. A
similar structure as for Al is evident. The possibility of experimental
observations of Ti \ka resonances at the LCLS-XFEL suggest itself, at
the averaged energies and strengths given in Table~1 for each Ti ion.
While for Al ions \ka resonance fluorescence occurs in the 1.48-1.88
keV range, the corresponding range for Ti \ka resonances is 4.5-5 keV.
The LCLS-XFEL is capable of energies up to about 8 keV in the
fundamental mode, well above the Ti \ka energies (or even the Fe
\ka energies $<$6.9 keV [1]). A
more complete account of the atomic calculations and models of related processes
will be given elsewhere. Here we note that in addition to the primary
physical process of an L-shell vacancy pumped by K-shell
photo-excitation, there are several, often competing, processes that
need to be accounted for. These include Auger decays and radiative cascades
from outer shells, electron impact ionization dependent on plasma
density, and resonance broadening and overlap.

\begin{table*}
\caption{Averaged $K\alpha$ Resonant Energies and Cross Sections for
Al and Ti Ions}
\begin{centering}
\begin{tabular}{l@{\hspace{.3em}}c@{\hspace{.3em}}c@{\hspace{.3em}}c@{\hspace{.3em}}c@{\hspace{.3em}}c@{\hspace{.3em}}c}
\hline
 & &  & \multicolumn{2}{c}{Al} & \multicolumn{2}{c}{Ti} \tabularnewline
\hline
%Ion & Transition Array & \# of \(K_{\alpha}\) & \(\langle
%E(K\alpha)\rangle\) & 
Ion & Transition Array & \# of Tran- & \(\langle E(K_\alpha)\rangle\) &
\(\langle \sigma_{res}(K_{\alpha})\rangle\) & \(\langle
E(K_\alpha)\rangle\) &
\(\langle \sigma_{res}(K_{\alpha})\rangle\) \tabularnewline
Core& & sitions & (keV) & (Mb) & (keV) & (Mb) \tabularnewline
\hline
F-like &
\(1s^{2}2s^{2}2p^{5}-1s2s^{2}2p^{6}\)&
2 &
1.491 &
1.40 & 4.541 & 1.73 \tabularnewline
O-like &
\(1s^{2}2s^{2}2p^{4}-1s2s^{2}2p^{5}\)&
14 &
1.509 &
7.05 & 4.575 & 8.63 \tabularnewline
N-like &
\(1s^{2}2s^{2}2p^{3}-1s2s^{2}2p^{4}\)&
35 &
1.522 &
11.2 & 4.604 & 13.3 \tabularnewline
C-like &
\(1s^{2}2s^{2}2p^{2}-1s2s^{2}2p^{3}\)&
35 &
1.539 &
15.4 & 4.639 & 17.9 \tabularnewline
B-like &
\(1s^{2}2s^{2}2p^{1}-1s2s^{2}2p^{2}\)&
14 &
1.556 &
8.20 & 4.671 & 9.22 \tabularnewline
Be-like &
\(1s^{2}2s^{2}-1s2s^{2}2p\)&
2 &
1.574 &
4.93 & 4.708 & 5.47 \tabularnewline
Li-like &
\(1s^{2}2s-1s2s2p\)&
6 &
1.587 &
5.70 & 4.732 & 6.01  \tabularnewline
He-like &
\(1s^{2}-1s2p\)&
2 &
1.599 &
6.11  & 4.755 & 6.24 \tabularnewline
H-like &
\(1s-2p\)&
2 &
1.789 &
3.33 & 4.975 & 3.28 \tabularnewline
\hline 
\end{tabular}
\par\end{centering}
\end{table*}

\begin{figure*}
%\begin{center}
%\includegraphics[width=160mm,height=95mm]{Kres.eps}
\includegraphics[width=160mm,height=95mm]{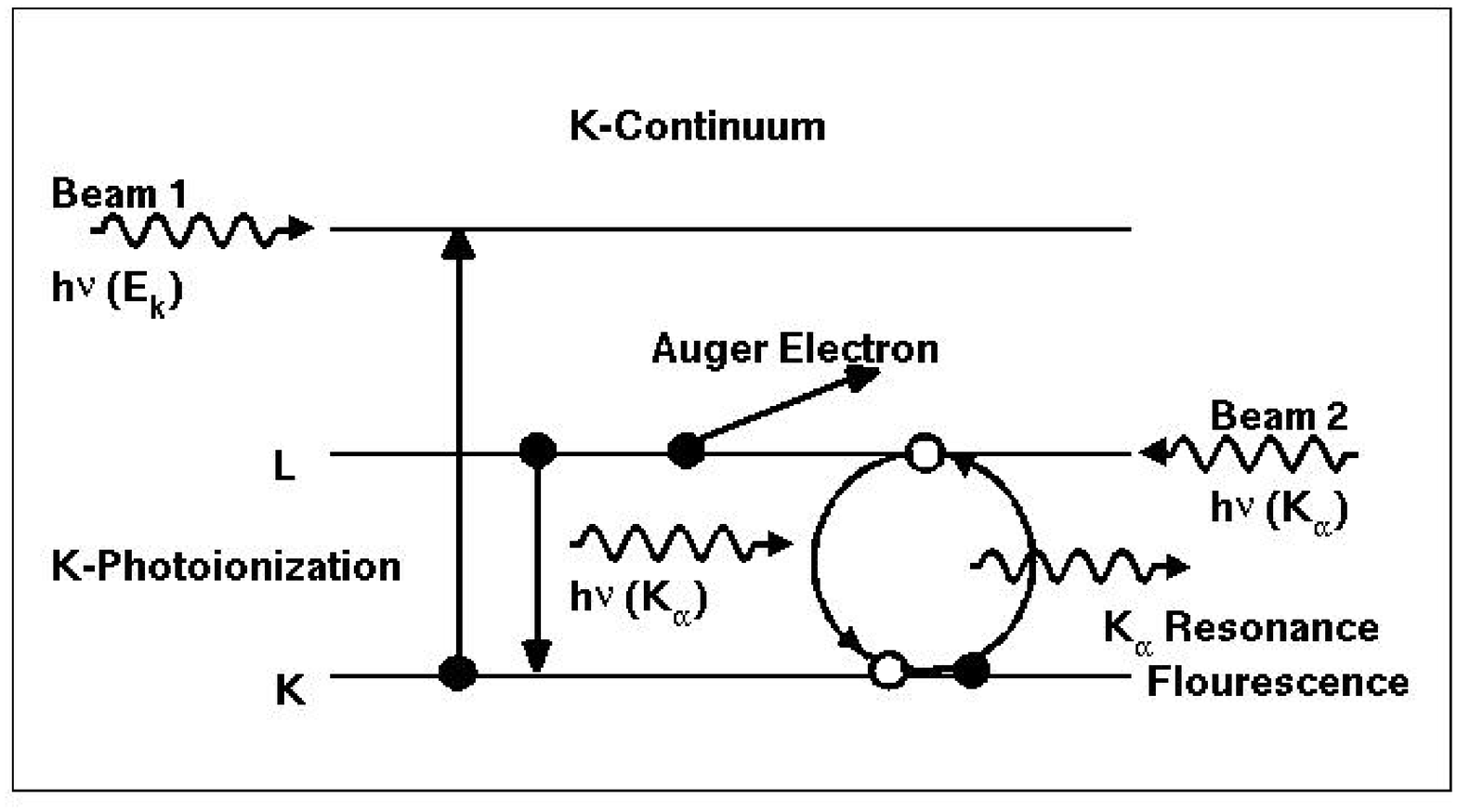}
%\end{center}
\caption{Schematic diagram of a twin-beam monochromatic X-ray system.
K-shell ionization by Beam 1 (on the left)
triggers the Auger decays, and \ka resonance 
fluorescence is pumped and driven by Beam 2 (on the right).
\label{fig:rfl}}
\end{figure*}

 The fact that a direct correspondence between \ka resonance complexes
and the \ka emission in a high-density and high-intensity environment
created in an XFEL can be established, reveals not only the nature
of these resonances but also their excitation mechanism. However, highly
intense monochromatic XFEL beams or synchrotron sources producing a HED plasma
are impractical for most common applications and
environments unable to withstand such intensities. 
Other monochromatic X-ray sources are being developed using peta-watt lasers
for plasma imaging and \ka radiography(e.g. \cite{park,akli11a,akli11b}). 
These are also very intense sources, with \ka conversion efficiencies of 
$\sim10^{-4}$ in the range of laser intensities of $10^{18-20}$ W/cm$^2$. 
On the other hand, a low intensity monochromatic X-ray source
may not contain sufficient fluence to be effective in creating and
pumping \ka RFL for useful purposes. In particular, monochromatic X-ray
biomedical imaging and therapeutics would need far less
intense fluxes.  One may therefore
ask the question (albeit theoretically): what is the minimum
intensity of an X-ray beam, or a combination
of x-ray beams, that can be employed in principle to excite or pump \ka
RFL? 
The femto-second, or
shorter, timescales are commensurate with atomic transition rates
(Einstein A and B-coefficients), and hence Auger decay radiative cascade 
rates. A monochromatic X-ray beam could
produce enough \ka photon flux to pump K-shell electrons, in
competition with downward decay rates, provided L-shell vacancies exist
{\it a priori}.  

 One possibility to minimize the incident flux required for \ka RFL 
is a twin-beam monochromatic X-ray device \cite{monox}, 
schematically illustrated in Fig.~\ref{fig:rfl}. The two beams are tuned
to the K-edge and the \ka resonance energies respectively. K-shell
ionizations induced by the first beam would lead to electron vacancies in
the L and higher shells via Auger decays. The second beam would pump the
RFL mechanism, as well as cause secondary ionizations from the emitted
photon and electron ejections. The excitation/ionization processes are
coupled, and dependent on photon fluences of the two beams.
Ultrafast monochromatic X-ray sources, such as the ones based on femtosecond PW
lasers, would be suitable for the twin-beam configuration. 
A generalization of the proposed embodiment in Fig.~\ref{fig:rfl}, 
utilizing multiple
monochromatic X-ray beams, may also be implemented. Such an extended
system could target higher than \ka resonance complexes, as discussed in
[1].
 
 Potential applications of monochromatic X-ray
systems have been proposed in biomedical spectroscopy for imaging and
therapy. For example, synchrotron sources tuned to the K-edge
have been employed to explore the breakdown of
high-Z compounds of platinum or gold in
nanomoities, to be deliverd to tumors and kill cancer cells
upon irradiation \cite{elleaume,lesech,yang,porcel}. 
Recent theoretical studies have shown
that relatively low energy X-rays in the E $<$ 100 keV range 
should be far more
effective than the conventional high energy X-rays
in the MeV range generated by linear accelerators (LINACs) used in
radiation therapy (e.g. \cite{leung,berbeco,jain,lim}). 
Monochromatic X-ray systems,
if available, might be ideal. The radiation dose
generally tolerated in radiation treatiment is approximately 1 Gray (Gy)
per minute (usually not to exceed a few tens of Gys). Given that 1 Gy =
1J/Kg, PW lasers, or combination thereof, can
generate this amount of X-ray flux for radiation treatment
(a radiation dose of 1 Gy is roughly equivalent to approximately 
10$^{10}$ photons of 80 keV energy directed at body tissue with area 1 cm$^2$).

A serious
problem with E$<$ 100 keV radiation is the factor of 2 or 3 larger attenuation
coefficients inside the body than the high energy MeV X-rays, rendering
the former unsuitable for treatment of deeply located tumors inside the
body; that is the {\it
raison d'etre} for employing high energy LINACs \cite{lim}. 
But the cross sections for photoionization, that trigger Auger decays and 
ejected electron yields, decrease as $\sigma_{PI} \sim
E^{-3}$, whereas Compton scattering cross sections increase with energy
until they exceed photoelectric absorption by orders of magnitude 
well below the MeV range \cite{nist}.
Therefore high energy photons are preferentially scattered inside the
body, rather than result in Auger electron yields that might
destroy cancer cells sensitized with high-Z nanomaterials.
Monte Carlo numerical simulations 
have shown that radiosensitization factors of Pt or Au with low energy
X-rays with mean energies E $\sim$ 100 keV are an order of magnitude higher, 
and therefore could more than
compensate for their reduced attenuation and be more effective than high
energy MeV photons \cite{lim}. Also, the use of broadband radiation
sources, such as the LINACs, is also wasteful since bremsstrahlung
output spectrum lacks specificity in energy and penetration depth.  
Therefore, tunable monochromatic X-ray sources would be preferable.
In addition, the efficacy of
high-Z contrast agents could be greatly enhanced if the \ka RFL
mechanism described in this {\it Letter} can be implemented in practice.
Driving the \ka Auger cycle with monochromatic X-ray system(s) would
result in increased local energy deposition than by K-shell ionization
alone. 

 A numerical collisional-radiative model may be constructed to simulate
the physical processes illustrated in Fig.~\ref{fig:rfl}. The model
would employ atomic rates for excitation, ionization, photon fluences
$\Phi_1, \Phi_2$ in beams 1 and two respectively, K and L level
populations, cascade coefficients from upper shells, etc. For instance,
the L-shell 
population for each ion is governed by direct photoionizations by the two
beams with $\Phi_1$ and $\Phi_2$, collisional ionizations by electrons
in the plasma at local density and temperature, cascades from outer
shells, resonant excitation from the K-shell as well as stimuated
emission L \ra K that constitutes the Auger cycle. Also, given the
photon fluences in the two beams, we may obtain an estimate of induced Rabi
oscillations at frequency $\omega_R = \mu_{KL} E/\hbar$, where
$\mu_{KL}$ is the diploe moment related to the A-coefficient for 
a given K \ra L transition, and $E$
is the electric field amplitude corresponding to the irradiance
(time-averaged power per unit area) $I = \frac{1}{2}c\epsilon_oE^2$  in
beam 2 with fluence $\Phi_2$. Though complex, such as model is
comuputationally feasible. However, the primary requirement is the
calculation of the cross sections, transition probabilities, and rates
for all contributing processes mentioned above. Work is in progress
along these directions.

 We would like to thank Sara Lim and Michael Dance for contributions. 
This work was
partially supported by grants from the U.S. Department of Energy,
National Nuclear Stewardship Alliance program (DE-FG52-09NA29580),
and the National Science Foundation (AST-0907763).

$\ast$ nahar.1@osu.edu, $\dagger$ pradhan.1@osu.edu

\end{document}